\documentclass[final,3p,times]{elsarticle}
 \biboptions{comma,sort&compress}
\usepackage{here}
 \usepackage{graphicx}
  \usepackage{epsfig}

\usepackage{amsmath,bm}
\usepackage{amssymb}
\usepackage[dvipsnames]{xcolor}

\def\beq{\begin{equation}}
\def\eeq{\end{equation}}
\def\bea{\begin{eqnarray}}
\def\eea{\end{eqnarray}}

\def\lesssim{\ \hbox{\raise 2pt \hbox{$<$} \kern -13pt
                     \lower 3pt \hbox{$\sim$}}\ }
\def\greatersim{\ \hbox{\raise 2pt \hbox{$>$} \kern -13pt
                     \lower 3pt \hbox{$\sim$}}\ }
\def\frac#1#2{ {{#1} \over {#2} }}

\input epsf.tex
\def\desepsf(#1 width #2){\epsfxsize=#2 \epsfbox{#1}}

\newcommand{\ot}{\leftarrow}

\renewcommand{\vec}[1]{\bm{#1}}

\def\({\left(}
\def\[{\left[}
\def\){\right)}
\def\]{\right]}

\newcommand{\specialcellcenter}[2][c]{\begin{tabular}[#1]{@{}c@{}}#2\end{tabular}}

\begin{document}

\begin{frontmatter}

\vspace*{1.4 cm} 
\title{Non-perturbative contributions to vector-boson transverse momentum spectra\\ in hadronic collisions}
\author[label1,label2]{Francesco Hautmann}
 \address[label1]{Rutherford Appleton Laboratory, Chilton OX11 0QX and Physics  Department, University of Oxford,  Oxford OX1 3NP, United Kingdom}
\address[label2]{Elementaire Deeltjes Fysica, Universiteit Antwerpen, B 2020 Antwerpen, Belgium}
 \author[label3]{Ignazio Scimemi}
\address[label3]{Departamento de F\'isica Te\'orica and IPARCOS, Universidad Complutense de Madrid (UCM), 28040 Madrid, Spain}  
\author[label4]{Alexey Vladimirov} 
\address[label4]{Institut f\"ur Theoretische Physik, Universit\"at Regensburg, D-93040 Regensburg, Germany}

\begin{abstract}
Experimental  measurements of Drell-Yan 
(DY) vector-boson production 
are available from the Large Hadron Collider (LHC) and from lower-energy collider and fixed-target experiments. In the 
region of low vector-boson transverse momenta   $q_T$, 
which is important  
for the extraction of the $W$-boson mass at the LHC, 
  QCD contributions from non-perturbative Sudakov form factors and 
intrinsic transverse momentum distributions 
become relevant.  
We study the potential for determining such 
 contributions from fits to LHC and lower-energy experimental 
 data, using the framework of low-$q_T$ factorization for DY differential 
 cross sections in terms of 
 transverse momentum dependent (TMD) distribution functions. We 
 investigate correlations between different sources of TMD 
non-perturbative effects, and correlations with collinear 
parton distributions. We stress the relevance of accurate 
 DY measurements at low masses and with fine binning in transverse momentum for improved determinations of long-distance contributions to Sudakov evolution 
processes and   TMDs. 
\end{abstract} 
\end{frontmatter}

\section{Introduction} 

The production of   photons, weak bosons  and leptons 
at large momentum transfer 
$Q \gg \Lambda_{\rm{QCD}}$ in high-energy hadronic collisions 
is described successfully by factorization~\cite{Collins:1989gx} of short-distance hard-scattering cross sections, computable at finite order in 
QCD perturbation theory as  power series expansions in the strong coupling $\alpha_s$, and non-perturbative long-distance parton distribution 
 functions  (PDFs), determined from fits to experiment. 
It was realized long ago, however, that 
even for  $Q \gg \Lambda_{\rm{QCD}}$ 
additional dynamical effects need to be taken into account to describe 
physical  spectra in the vector-boson 
transverse momentum $q_T$ when  the multiple-scale 
region  $q_T \ll Q$ is 
reached~\cite{Dokshitzer:1978hw,Parisi:1979se,Curci:1979bg,Collins:1981uk}. 
These amount to 
  i)    perturbative logarithmically-enhanced corrections in $ \alpha_s^k  \ln^m Q / q_T$ ($m \leq 2k$), which go  
  beyond finite-order  perturbation theory and call for  summation  to all   
  orders  in $\alpha_s$, and ii) non-perturbative contributions  besides PDFs,   
 which  correspond to the intrinsic 
transverse momentum distributions  in the initial states of the collision 
 and to non-perturbative components of  
Sudakov form factors. 

The summation of the logarithmically-enhanced corrections to  Drell-Yan (DY) lepton pair hadroproduction 
has since been accomplished systematically 
by methods based on the CSS formalism~\cite{Collins:1984kg}. It has been fully computed 
through next-to-next-to-leading-logarithmic  (NNLL)   accuracy, which requires calculations up to two-loop level, and 
partial results at three and four loops are already available for some of the coefficients needed for higher logarithmic 
accuracy~\cite{Vogt:2018miu,Luo:2019szz}.     
On the other hand,  nonperturbative effects besides  PDFs in DY production are included in the formalism of 
transverse momentum dependent (TMD)  parton distribution   
functions~\cite{Angeles-Martinez:2015sea}. 
Intrinsic transverse momentum distributions enter as  boundary conditions  to the renormalization 
group evolution equations for TMDs, while  non-perturbative Sudakov effects 
 are taken into account via 
non-perturbative contributions to the kernel of the evolution equations 
associated with TMD rapidity 
divergences~\cite{Collins:1999dz,Collins:2003fm,Hautmann:2007uw,Collins:2011zzd,GarciaEchevarria:2011rb}.   

The purpose of this  work is  to examine   the combined determination of   the  nonperturbative 
rapidity-evolution kernel  and intrinsic transverse momentum $k_T$ distribution  from   fits to  
 measurements of  transverse momentum spectra  in DY lepton-pair production at the Large Hadron Collider (LHC) and in 
 lower-energy  experiments, including Tevatron, RHIC and fixed-target experiments. To this end, we  
  employ the calculational  framework developed in~\cite{Scimemi:2017etj,Scimemi:2018xaf,Bertone:2019nxa,Vladimirov:2019bfa,Scimemi:2016ffw,Scimemi:2019cmh}.  
We investigate to what extent   the two  sources of non-perturbative effects  
are correlated,  and 
study the role of different data sets, from the high-precision DY LHC data to the lower-energy DY data, in disentangling them.  We also analyze how these 
two non-perturbative contributions are correlated with non-perturbative contributions encoded in  PDF sets. 
Quantifying these effects will be important both for strong interaction investigations of 
hadron structure and for determinations of precision electroweak parameters, as 
the  low-$q_T$  DY  region is relevant  for 
 the extraction of the $W$-boson mass at the LHC. 

The paper is organized as follows. In Sec.~2 we briefly describe the factorization formula,  evolution equations and perturbative coefficients which constitute the 
theoretical inputs to our analysis.   In Sec.~3 we present the results of the  numerical studies and fits to experimental data.  We give conclusions in Sec.~4.

\section{Theoretical inputs} 

We start from the TMD factorization formula for the differential cross section for DY lepton pair production 
 $h_1+h_2\to Z/\gamma^*(\to ll')+X$ at low  $ q_T \ll Q$~\cite{Collins:2011zzd}  
\begin{eqnarray}\label{def:xSec}
\frac{d\sigma}{dQ^2 dy dq_T^2}=\sigma_0\sum_{f_1,f_2}H_{f_1f_2}(Q,\mu)\int \frac{d^2\vec b}{4\pi} 
e^{i \vec b\cdot \vec q_T }F_{f_1\ot h_1}(x_1,\vec b;\mu,\zeta_1)F_{f_2\ot h_2}(x_2,\vec b;\mu,\zeta_2) 
+ {\cal O} \left( {q_T / Q} \right) + {\cal O} \left( {\Lambda_{\rm{QCD}} / Q} \right) ,
\end{eqnarray}
where $Q^2$, $\vec q_T$ and $y$ are the invariant mass, transverse momentum and rapidity of the lepton pair,  
 and  the TMD distributions  $F_{f\ot h}$  fulfill   evolution equations   in rapidity 
\begin{equation} 
\label{rapevolF} 
 {{\partial \ln F_{f\ot h} }  \over {\partial \ln \zeta} }    = -   \mathcal{D}^f(\mu,\vec b)    
\end{equation} 
and in mass 
\begin{equation} 
\label{mass-evol} 
  {{ \partial  \ln  F_{f\ot h}  }  \over {\partial  \ln \mu} }   =  \gamma_F  (\alpha_s (\mu) , \zeta / \mu^2 )  \;\; , 
  \;\;\;\;    {{ \partial  \mathcal{D}^f(\mu,\vec b)     }  \over {\partial  \ln \mu} }   =   {1 \over 2} \ 
  \Gamma_{\rm{cusp}}  (\alpha_s (\mu)  )  \;\; . 
\end{equation} 
We further perform the small-$\pmb b$  operator product expansion of 
the TMD  $F_{f\ot h}$ as follows,     
\begin{eqnarray}\label{model:TMDPDF}
F_{f\to h}(x,\vec b)=f_{\rm NP}(x,\vec b)\sum_{f'}\int_x^1\frac{dy}{y}C_{f\ot f'}\(\frac{x}{y},\ln\(\vec b^2 \mu^2\)\)f_{f'\ot h}(y,\mu),
\end{eqnarray}
where  $f_{f'\ot h}$ are the PDFs,  $C_{f\ot f'}$ are the matching Wilson coefficients,  and    $f_{\rm NP}$ are 
functions\footnote{In full generality,   the functions $f_{\rm NP}$ may depend on  flavor and on the convolution variable $y$. We do not consider these more 
general scenarios here.}  to be fitted to data, encoding non-perturbative information about  the intrinsic transverse momentum distributions. 
The non-perturbative component  of the rapidity-evolution kernel  $  \mathcal{D}^f $ and the distribution    $f_{\rm NP}$  are the main focus of this paper. 
 
The  TMD distributions in Eq.~(\ref{def:xSec}) depend on the scales   $\mu,\zeta$.   To set  these scales, we will use the method of the 
$\zeta$-prescription proposed in~\cite{Scimemi:2017etj}. (See e.g.~\cite{Bacchetta:2019sam} for recent examples of alternative scale-setting.) 
The summation of the logarithmically-enhanced corrections 
at low  $ q_T$ is achieved through Eqs.~(\ref{def:xSec})-(\ref{model:TMDPDF}) 
by computing perturbatively 
the quantities $H$, $C$, $\gamma_F$ and $\Gamma_{\text{cusp}}$ 
  as  series expansions in powers of $\alpha_s$. 
In Table~\ref{tab:pert} we summarize the perturbative  orders 
 used for each of these quantities in the calculations that 
 follow. We refer to the logarithmic accuracy specified by these orders as NNLL.\footnote{Different terminologies 
 are also in use in the literature (see e.g.~\cite{Bacchetta:2019sam}). For instance,  $H$ elements of Table~\ref{tab:pert}  
 are sometimes referred to 
 as NNLL$^\prime$, and $\gamma_F$ elements as  N$^3$LL.}  

\begin{table}[h]
\begin{center}
\begin{tabular}[h]{||c|c||c|c||c|c||}\hline
  $H$ &  $C_{f\ot f'}$ & $\Gamma_{\text{cusp}}$ & $\gamma_F$ & $\alpha_s$ running & PDF evolution
\\\hline 
  $\alpha_s^2$ & $\alpha_s^2$ & $\alpha_s^3$  & $\alpha_s^3$ & \multicolumn{2}{|c||}{\specialcellcenter{NNLO 
} }
\\\hline
\end{tabular}
\end{center}
\caption{\label{tab:pert} Summary of perturbative orders used  for each part of the DY cross section.}
\end{table}

The rapidity evolution kernel  $\mathcal{D}$ contains perturbative and nonperturbative components. 
 The perturbative expansion   for 
 $\mathcal{D}$ is currently known up to three-loops~\cite{Moch:2005tm,Baikov:2009bg,Vladimirov:2016dll,Li:2016ctv}. 
Using the $b^*$ prescription~\cite{Collins:1984kg}, we model  $\mathcal{D}$   as
\begin{eqnarray}\label{model:rad}
\mathcal{D}^f(\mu,\vec b)=\mathcal{D}^f_{\text{res}}\(\mu,b^*(\vec b)\)+g(\vec b),
\end{eqnarray}
where $\mathcal{D}^f_\text{res}$~\cite{Echevarria:2012pw} is the resummed perturbative part of $\mathcal{D}^f$, 
$g$ is an even function of $\vec b$ vanishing as $\vec b\to 0$, and
\begin{eqnarray}\label{eq:bstar}
b^*(\vec b)=|\vec b|\sqrt{\frac{ B_{\text{NP}}^2}{\vec b^2+B_{\text{NP}}^2}}\, ,   
\end{eqnarray}
with the parameter $ B_{\text{NP}} $  to be fitted to experimental data. 
For the function $ g(\vec b)  $ we will use the models 
\begin{eqnarray}\label{model:exp-gk}
g(\vec b)=g_K \vec b^2  , 
\end{eqnarray}
\begin{eqnarray}\label{model:exp}
g(\vec b)=c_0 |\vec b| b^*(\vec b),
\end{eqnarray}
and 
\begin{eqnarray}\label{model:exp-gk*}
g(\vec b)=g_K^* \vec b^{*2} , 
\end{eqnarray} 
fitting respectively the parameters $ g_K$,  $c_0$  and $ g_K^*  $   to experimental data.  
The quadratic model in  Eq.~(\ref{model:exp-gk})  has traditionally been 
used since the pioneering works~\cite{Ladinsky:1993zn,Landry:1999an,Landry:2002ix,Konychev:2005iy}. 
The model in Eq.~(\ref{model:exp-gk*})   contains 
 the perturbative quadratic behavior  at small  $ |\vec b| $  but it goes to a   
 constant behavior  at large $ |\vec b| $, fulfilling the asymptotic condition $ \partial  \mathcal{D} / \partial \ln \vec b^2 = 0 $,  
in a similar spirit to parton saturation in the $s$-channel picture~\cite{Hautmann:2007cx} for parton distribution functions. 
The model in  Eq.~(\ref{model:exp}) is   an 
intermediate model between the previous two, being 
characterized by a linear rise at large $ |\vec b| $. 
In the following we will refer to the 
non-perturbative component  of the rapidity-evolution kernel, modeled according to Eqs.~(\ref{model:exp-gk})-(\ref{model:exp-gk*}),  
as $D_{\rm NP}$. 

The nonperturbative contribution to $\mathcal{D}^f$  in   
Eq.~(\ref{model:rad})  also 
influences  the rapidity scale fixing with 
the $\zeta$-prescription~\cite{Vladimirov:2019bfa}.
In fact, once the nonperturbative correction is included  in $\mathcal{D}^f$,  
one is to use $\zeta_{NP}$ given by~\cite{Vladimirov:2019bfa}
\begin{eqnarray}\label{th:zetaNP}
\zeta_{\text{NP}}(\mu,b)=\zeta_{\text{pert}}(\mu,b)e^{-\vec b^2/B_{\text{NP}}^2}+\zeta_{\text{exact}}(\mu,b)\(1-e^{-\vec b^2/B_{\text{NP}}^2}\).
\end{eqnarray}
 Only the perturbative part $\zeta_{\text{pert}}$, computed in~\cite{Scimemi:2018xaf},   
was used in the fits~\cite{Bertone:2019nxa}. The  expression  in Eq.~(\ref{th:zetaNP}) converges to $\zeta_{\text{pert}}$ in the limit $b\to 0$.
We will use this expression in the fits of the next section. 

The modeling of the TMD through the function $f_{\rm NP}$ allows one to fit  data at 
 different energies. In particular it allows  the nonperturbative behavior of the TMD to be described for large values of  $b$.
In~\cite{DAlesio:2014mrz,Scimemi:2017etj,Bertone:2019nxa} it has been observed that a modulation between 
Gaussian and exponential models  is necessary. This can be provided by the following model, 
\begin{eqnarray}\label{model:our_fNP}
f_{\rm NP}(x,\vec b)=\exp\(-\frac{(\lambda_1(1-x)+\lambda_2 x+\lambda_3 x(1-x))\vec b^2}{\sqrt{1+\lambda_4 x^{\lambda_5} \vec b^2}}\),
\end{eqnarray}
where the interpolation of Gaussian/exponential regimes is dependent on the Bjorken $x$-variable,   
and $\lambda_{1,..,5}>0$.

\section{Determination of $f_{\rm{NP}}$  and $D_{\rm{NP}}$ from fits to experiment}

We  next present  results of performing TMD fits to experimental data for  
DY differential cross sections, by employing 
 the theoretical framework described in the 
previous section. We consider DY measurements both 
at the LHC~\cite{Aad:2014xaa,Aad:2015auj,Chatrchyan:2011wt,Khachatryan:2016nbe,Aaij:2015gna,Aaij:2015zlq,Aaij:2016mgv} and in lower-energy 
experiments~\cite{Aidala:2018ajl,Aaltonen:2012fi,Affolder:1999jh,Abazov:2010kn,Abazov:2007ac,Abbott:1999wk,McGaughey:1994dx,Moreno:1990sf,Ito:1980ev}. 
The fits are performed  using the code \texttt{artemide}~\cite{web,Scimemi:2017etj}.   
A detailed technical description of the methodology used for  
 these fits is reported in~\cite{Scimemi:2019cmh}.   
 
\begin{figure}[h]
\begin{center}
\includegraphics[width =0.45\textwidth]{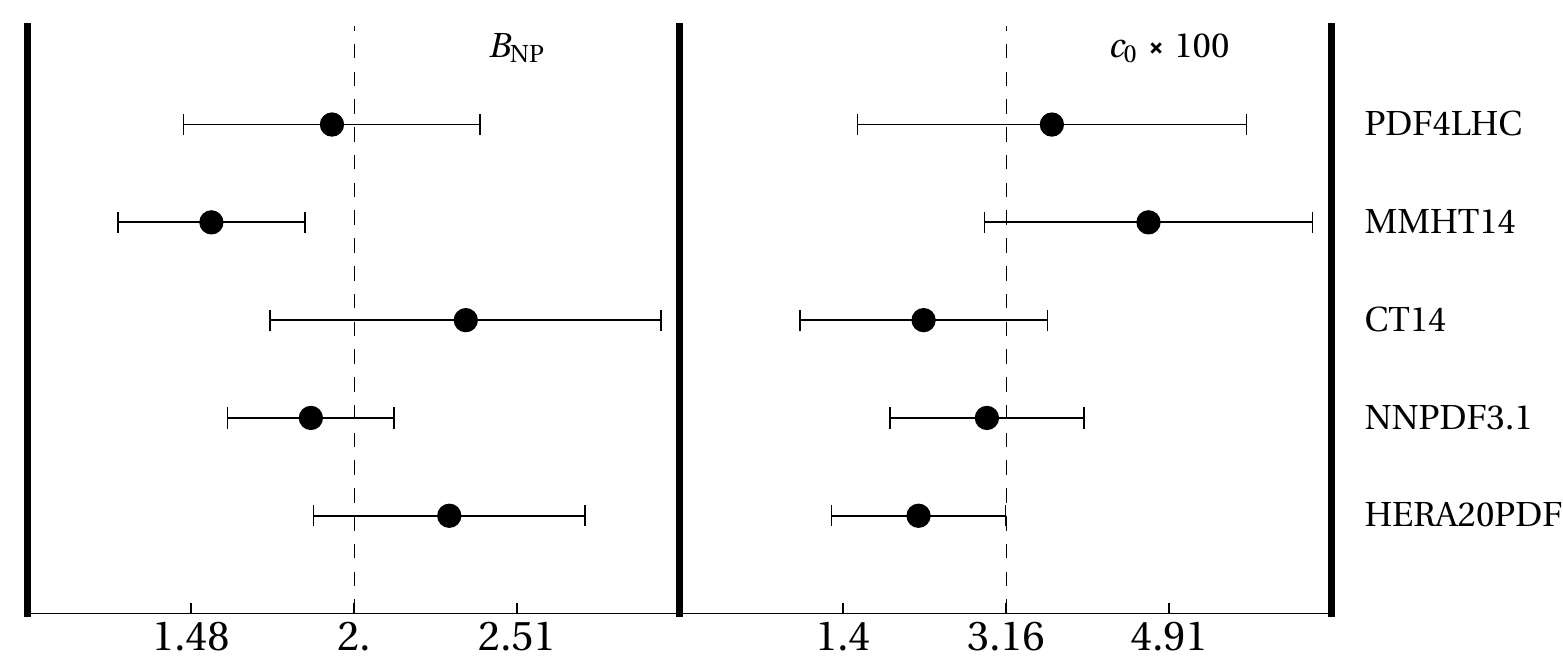}\\
\includegraphics[width =1\textwidth]{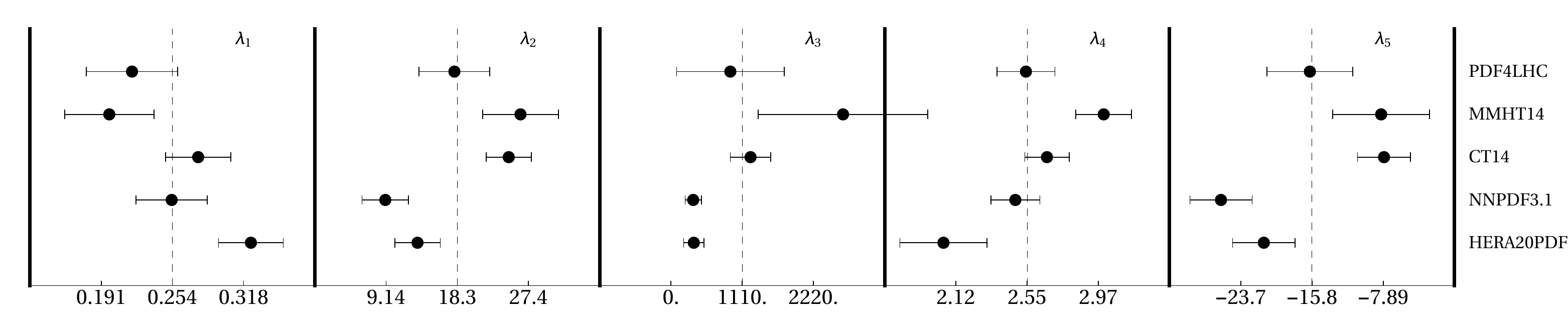}
\end{center}
\caption{Results of the TMD global fit to DY measurements from  
LHC and lower-energy experiments.}
\label{fig:PDFs}
\end{figure}

Let us start with  the global fit of the TMD parameters to DY 
LHC~\cite{Aad:2014xaa,Aad:2015auj,Chatrchyan:2011wt,Khachatryan:2016nbe,Aaij:2015gna,Aaij:2015zlq,Aaij:2016mgv} and 
lower-energy~\cite{Aidala:2018ajl,Aaltonen:2012fi,Affolder:1999jh,Abazov:2010kn,Abazov:2007ac,Abbott:1999wk,McGaughey:1994dx,Moreno:1990sf,Ito:1980ev} 
data.\footnote{Besides DY data, semi-inclusive deep inelastic scattering (SIDIS) measurements (see e.g.~\cite{Aghasyan:2017ctw,Airapetian:2012ki}) 
also provide powerful constraints on TMD parton distributions. In the case of SIDIS, however,   additional nonperturbative effects enter through TMD fragmentation 
functions. In the present paper we limit ourselves to TMD fits based on DY processes. See e.g.~\cite{Scimemi:2019cmh} for fits to both DY and SIDIS data.} 
We  restrict the fit  to data in the low transverse momentum region by  
 applying the cut $q_T/Q 
 < 
 0.2$ to the data sets.\footnote{In order to treat  the region  
$ q_T \sim Q$,   the matching of TMD contributions with finite-order NLO (or NNLO) perturbation theory becomes 
essential~\cite{Camarda:2019zyx,Catani:2015vma,Bizon:2019zgf,Bizon:2018foh,Martinez:2019mwt,Martinez:2018jxt}.  
See in particular the recent studies~\cite{Martinez:2020fzs,Bacchetta:2019tcu} of   the region of moderate transverse 
momenta and masses, using different 
matching methods~\cite{Collins:1984kg,Collins:2000gd}.}  
 The values of 
the fitted TMD parameters 
in Eqs.~(\ref{eq:bstar}),(\ref{model:exp}) (for $D_{\rm NP}$) and 
in Eq.~(\ref{model:our_fNP}) (for $f_{\rm NP}$)  
and their  associated uncertainties are shown in  Fig.~\ref{fig:PDFs}.  Since 
PDFs enter the TMD fit through Eq.~(\ref{model:TMDPDF}), the results 
in   Fig.~\ref{fig:PDFs} are presented for different PDF sets. The corresponding 
$\chi^2$ values are given in Table~\ref{tab:result-new}. 
We observe that the values of the  fitted parameters $\lambda_i$ 
(see Eq.~(\ref{model:our_fNP}))    
in   Fig.~\ref{fig:PDFs} vary more significantly 
among different PDF sets than the values of the fitted parameters 
$B_{NP}$ and  $c_0$ (see Eqs.~(\ref{eq:bstar}),(\ref{model:exp})),  
 corresponding to the fact that the $\lambda_i$ parameters in $f_{\rm NP}$ 
are related to  the $x$-dependence of the distributions, while the 
rapidity evolution kernel is $x$-independent. 

\begin{table}
\begin{center}
\begin{tabular}{||c|c||}
\hline
PDF & $\chi^2/$d.o.f.
 \\
\hline
NNPDF3.1~\cite{Ball:2017nwa} & 1.14\\
HERAPDF2.0~\cite{Abramowicz:2015mha} &0.97\\
CT14~\cite{Dulat:2015mca} &1.59\\
MMHT14\cite{Harland-Lang:2015nxa}&1.34\\
PDF4LHC~\cite{Butterworth:2015oua} &1.53\\
\hline
\end{tabular}
\caption{  PDF sets and $\chi^2$/d.o.f. results in a 
 TMD global fit to DY measurements.}
\label{tab:result-new}
\end{center}
\end{table}

\begin{figure}[h]
\begin{center}
\includegraphics[width =0.25\textwidth]{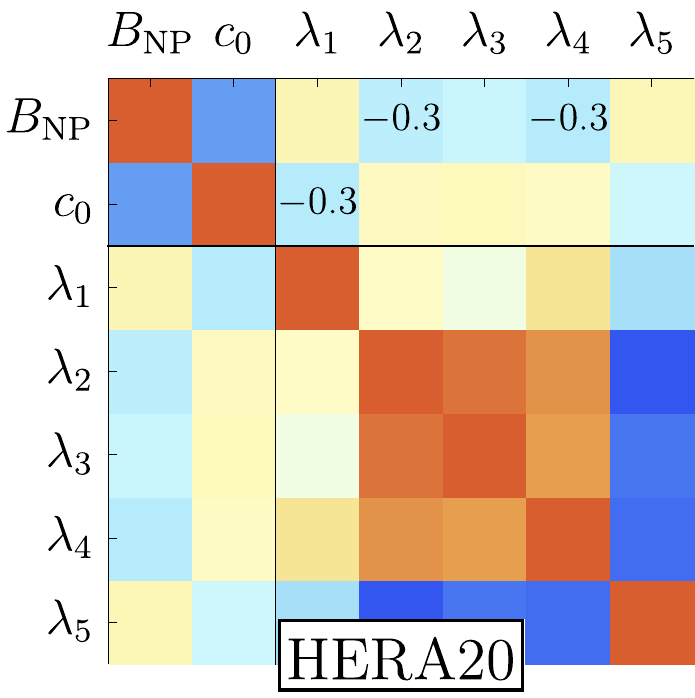}
\hspace{0.5cm}
\includegraphics[width =0.25\textwidth]{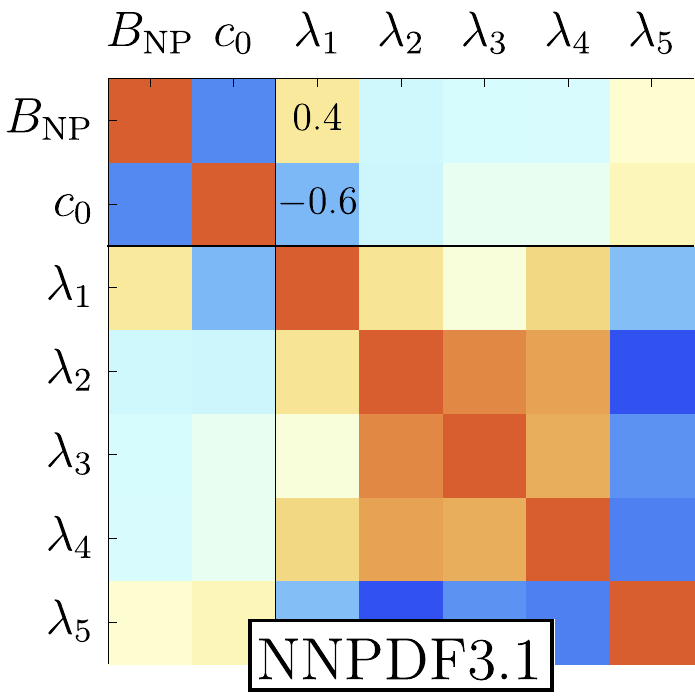}
\hspace{0.5cm}
\includegraphics[width =0.04\textwidth]{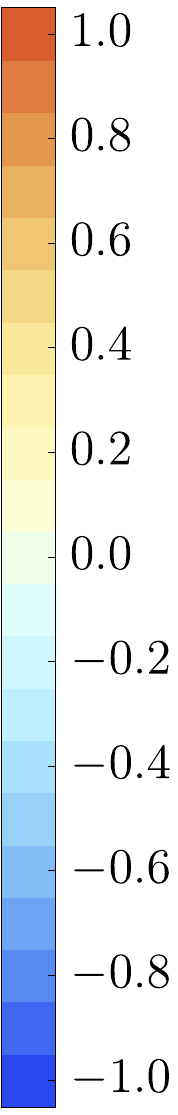}
\\
\includegraphics[width =0.25\textwidth]{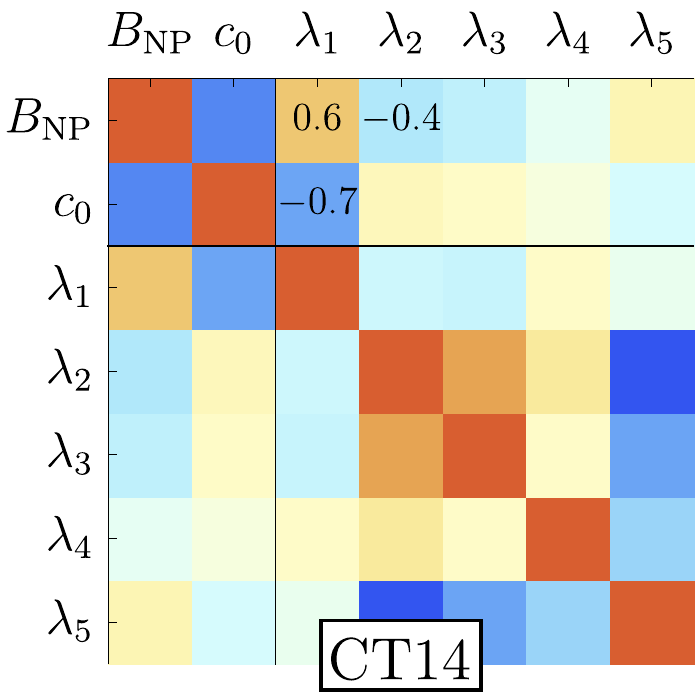}
\hspace{0.5cm}
\includegraphics[width =0.25\textwidth]{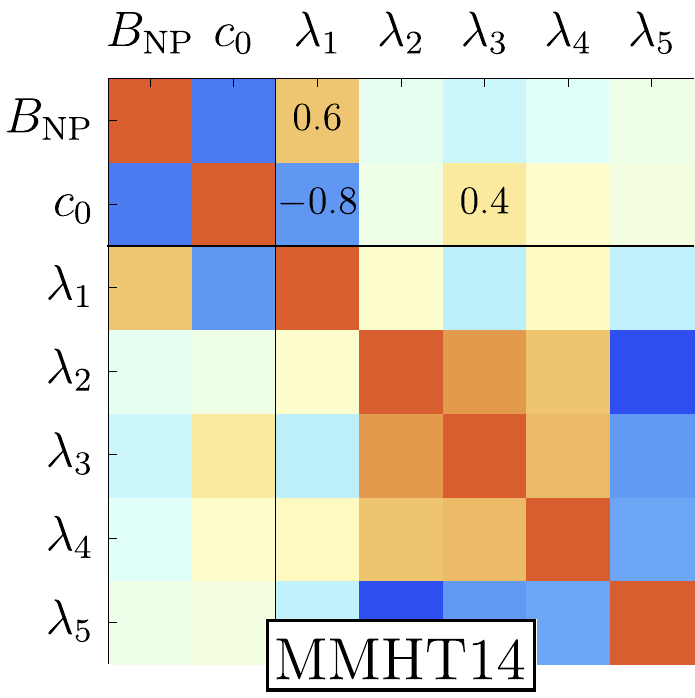}
\hspace{0.5cm}
\includegraphics[width =0.25\textwidth]{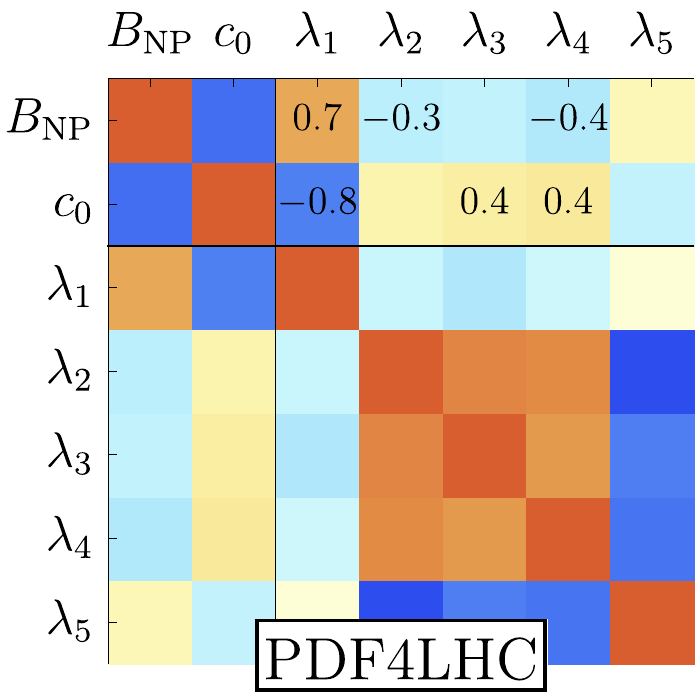}
\hspace{0.5cm}
\end{center}
\caption{Correlations of TMD fit parameters. In the axes $1=B_{NP}$, $2=c_0$, $(3,4,5,6,7)=\lambda_{1,2,3,4,5}$. Low correlation is represented by light colors, 
high correlation by dark colors. (The diagonal entries are trivial.) 
\label{fig:correlations}}
\end{figure}

The correlations among TMD parameters 
for different PDF sets are illustrated in Fig.~\ref{fig:correlations}. 
Light colors in the pictures of Fig.~\ref{fig:correlations} 
indicate low correlations; dark colors  indicate 
high correlations. Shades of blue 
denote negative correlations; shades of brown denote positive correlations. 
In particular,  the correlation between the parameters 
$c_0$ (controlling the long-distance behavior of the rapidity evolution kernel 
in Eq.~(\ref{model:exp})) and $\lambda_1$ (controlling the intrinsic transverse 
momentum distribution in Eq.~(\ref{model:our_fNP})) is  fairly low 
in the case of the HERAPDF set, but it increases in the NNPDF3.1 case, and is 
higher still in the CT14 and MMHT14 cases. 
We note that the latter two PDF sets  
do not include LHC data in the fits, while the NNPDF3.1 does. 
The $\chi^2$  values in Table~\ref{tab:result-new} are 
lowest for the HERAPDF and NNPDF3.1 cases. 

Next,  we wish to focus  on the role  
 of present (and future) LHC measurements  to investigate the  sensitivity 
to the nonperturbative contributions in $D_{\rm NP}$   and 
$f_{\rm NP}$. To this end   we will  perform fits 
to LHC data only,  using a smaller number of parameters. 
That is, we model  $D_{\rm NP}$  as in 
Eqs.~(\ref{model:rad})-(\ref{model:exp-gk*}), depending on 
two parameters,  $B_{NP}$ and either $g_K$ or $c_0$ or $ g_K^*  $,  
and we take  a form   for   $f_{\rm NP}$ which is 
simplified with respect to   Eq.~(\ref{model:our_fNP}), namely, 
we take an $x$-independent simple gaussian 
depending on one parameter $\lambda_1$ only,  which provides a measure of 
the intrinsic transverse momentum 
in terms of a gaussian width.  We then  perform   
3-parameter fits to LHC DY data~\cite{Aad:2014xaa,Aad:2015auj,Chatrchyan:2011wt,Khachatryan:2016nbe,Aaij:2015gna,Aaij:2015zlq,Aaij:2016mgv}, 
fitting $\lambda_1$, $B_{NP}$ and either $g_K$ or $c_0$ or $ g_K^*  $, 
 as well as  2-parameter fits to the same data,  
fitting only $B_{NP}$ and either $g_K$ or $c_0$ or $ g_K^*  $, and 
fixing $\lambda_1$ to 
$\lambda_1 = 0.001 $ GeV$^{2}$ 
to simulate the 
 cases  
of nearly zero intrinsic transverse momentum (as in 
purely collinear approaches).  
The results from the 3-parameter and 2-parameter fits, using 
the PDF set NNPDF3.1,    are summarized in Table~\ref{tab:case1}. 

\begin{table}[h]
\begin{center}
\begin{tabular}{||c|c|c|c|c|c||}
\hline\hline
Case  & $B_{NP}$& $g_K$ & $\lambda_1$  $(f_{NP}=\exp{-\lambda_1 b^2})$& $\chi^2/dof$ & $\chi^2/dof$(norm.)
\\ \hline
1     & 5.5 (max) & $0.116\pm 0.002$   & $10^{-3}$(fixed) & 3.29 & 3.04
\\ \hline
2      & $2.2\pm 0.4$ & $0.032\pm 0.006$   & $0.29\pm0.02$ & 1.50 &  1.28
\\ \hline
Case  & $B_{NP}$& $c_0$ & $\lambda_1$ & $\chi^2/dof$ & $\chi^2/dof$(norm.)
\\ \hline
3      & 1. (min) & $0.016\pm 0.001$   & $10^{-3}$(fixed) & 2.21 &  1.99
\\ \hline
4     & $3.0\pm 1.5$ & $0.04\pm 0.02$   & $0.27\pm0.04$ & 1.61 &  1.36
\\ \hline
Case  & $B_{NP}$& $g_K^*$ & $\lambda_1$ & $\chi^2/dof$ & $\chi^2/dof$(norm.)
\\ \hline
5      & $1.34\pm0.01$ & $0.16\pm 0.01$   & $10^{-3}$(fixed) & 1.70 & 1.52
\\ \hline
6 & $2.43\pm 0.66$ & $0.05\pm 0.02$ & $0.24\pm 0.04 $ & 1.49 & 1.28
\\
\hline
\end{tabular}
\caption{Results of 3-parameter and 2-parameter fits.  
The PDF set used is NNPDF3.1~\cite{Ball:2017nwa}.}
\label{tab:case1}
\end{center}
\end{table}

We see that the 3-parameter fits (cases 2, 4 and 6 in Table~\ref{tab:case1}) yield results,  
both for the $\chi^2$  values and for the values of the fitted TMD parameters,  
which are not dissimilar from the global fit results given earlier, supporting the 
overall consistency of the TMD picture of low-energy and high-energy DY data. 
These three cases correspond to the three different long-distance behaviors of 
 the rapidity-evolution kernel $  \mathcal{D}(\mu,\vec b)  $ in   Eqs.~(\ref{model:exp-gk})-(\ref{model:exp-gk*}). 
Case 2 and case 6, in particular, while giving fits of comparable quality, correspond to very different 
physical pictures of the nonperturbative component of $  \cal{D}$.  Case 2 extends the quadratic 
behavior to large distance scales (see Eq.~(\ref{model:exp-gk})). In contrast, case 6  
fulfills  the saturating condition $ \partial  \mathcal{D} / \partial \ln \vec b^2 = 0 $  at large $| \vec b |$ (see Eq.~(\ref{model:exp-gk*})).  
This is, to our knowledge,  the first time  that a full fit to low-$ q_T$  DY data  
is performed in the hypothesis  of long-distance saturating behavior of  
the rapidity-evolution kernel. 
 
The 2-parameter fits (cases 1, 3 and 5  
in Table~\ref{tab:case1}), on the other hand, show significantly different behaviors, 
characterized by somewhat higher $\chi^2$  values and especially by 
significantly different values of the $D_{\rm NP}$ fitted parameters.  This 
indicates that, although most of the sensitivity to the intrinsic transverse 
momentum distribution comes from the lower-energy measurements, 
non-negligible $f_{\rm NP}$ effects are present at the LHC too. In 
particular, Table~\ref{tab:case1} suggests that without any intrinsic transverse momentum 
distribution it may  be possible to describe DY data at the LHC but this would lead to a 
different determination  for  $B_{NP}$ and   the 
rapidity evolution kernel.  That is, intrinsic 
transverse momentum effects may be reabsorbed by changes in the $D_{\rm NP}$ fit.   

\begin{figure}[t]
\begin{center}
\includegraphics[width =0.3\textwidth]{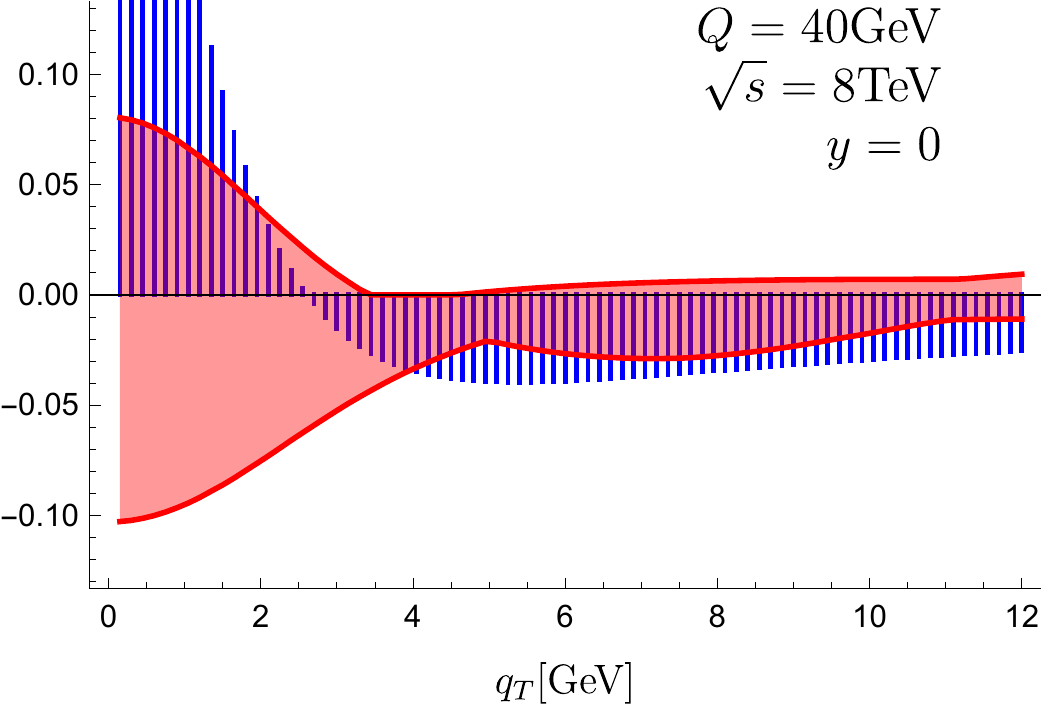}
\includegraphics[width =0.3\textwidth]{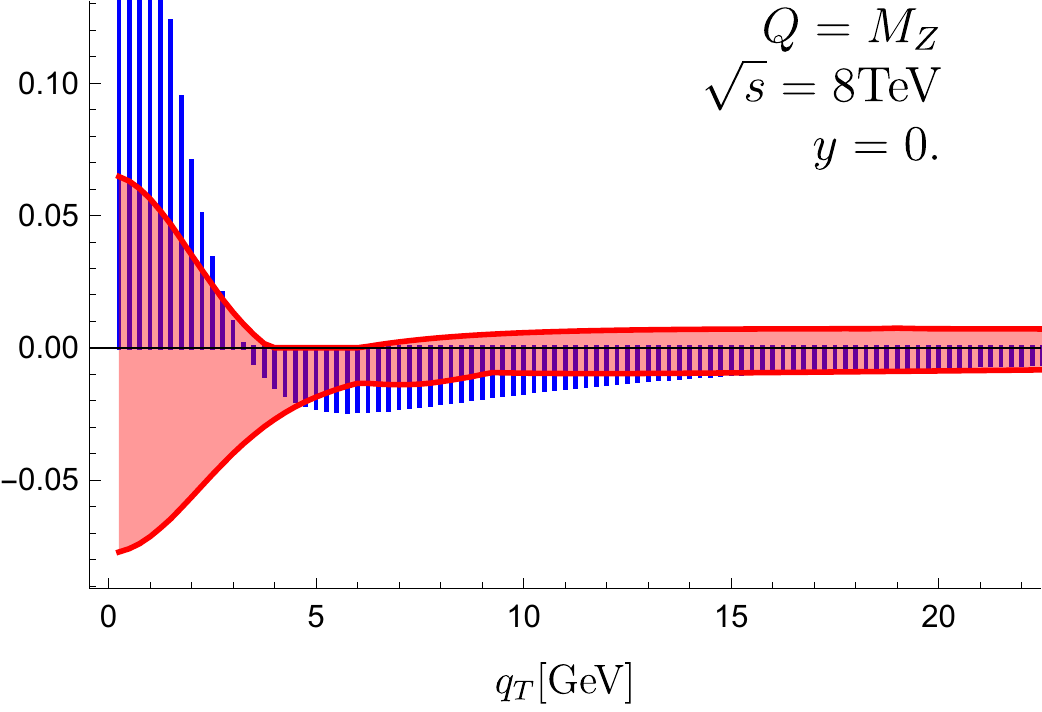}
\caption{Sensitivity to nonperturbative physics in LHC DY measurements: the transverse momentum dependence of the ratio 
in  Eq.~(\ref{eq:rsigma}),  for central rapidity and different values of the lepton-pair invariant mass. The solid band is 
obtained from perturbative scale variation.}  
\label{fig:NPs}
\end{center}
\end{figure}

To further analyze the sensitivity of LHC DY measurements to  $f_{\rm NP}$ and gain 
insight into the results of Table~\ref{tab:case1}, we next consider the ratio 
\begin{align}\label{eq:rsigma}
R_\sigma=2 \ \frac{d \sigma^{\rm test}-d \sigma^{TMD}}{d \sigma^{\rm test}+d \sigma^{TMD}} 
\  ,
\end{align}
where $ d \sigma^{TMD} $ is the DY differential cross section computed from the 
full TMD fit, and $ d \sigma^{\rm test} $ is the DY differential cross section computed 
by setting $f_{\rm NP}=1$ in the full fit. 
 In Fig.~\ref{fig:NPs} we plot the numerical results for the ratio (\ref{eq:rsigma})  
versus the DY lepton-pair transverse momentum $q_T$  for different values of 
the DY lepton-pair invariant mass $Q$. For reference, in  Fig.~\ref{fig:NPs}    
 we also plot the theoretical uncertainty band on the full TMD result which comes 
from scale variation, taken according to the $\zeta$ prescription of Sec.~2. We 
see that  in the lowest $q_T$ bins the nonperturbative effects,  evaluated 
according to the ratio in Eq.~(\ref{eq:rsigma}),   exceed the perturbative uncertainty,  
evaluated from scale variation in the $\zeta$ prescription.  The 
comparison of Table~\ref{tab:case1} and Fig.~\ref{fig:NPs} confirms  that 
sensitivity to $f_{\rm NP}$    is present in LHC 
data but may be reabsorbed by varying $D_{\rm NP}$.

\begin{figure}[t]
\begin{center}
\includegraphics[width =0.3\textwidth]{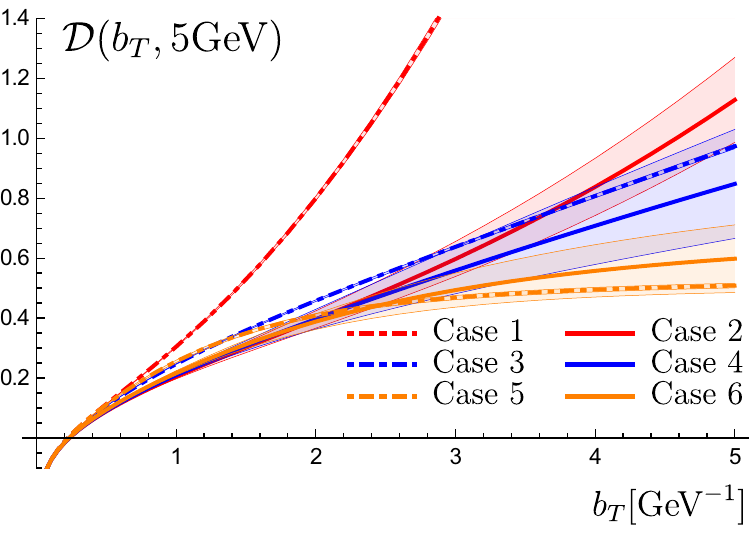}
\hspace{2cm}
\includegraphics[width =0.3\textwidth]{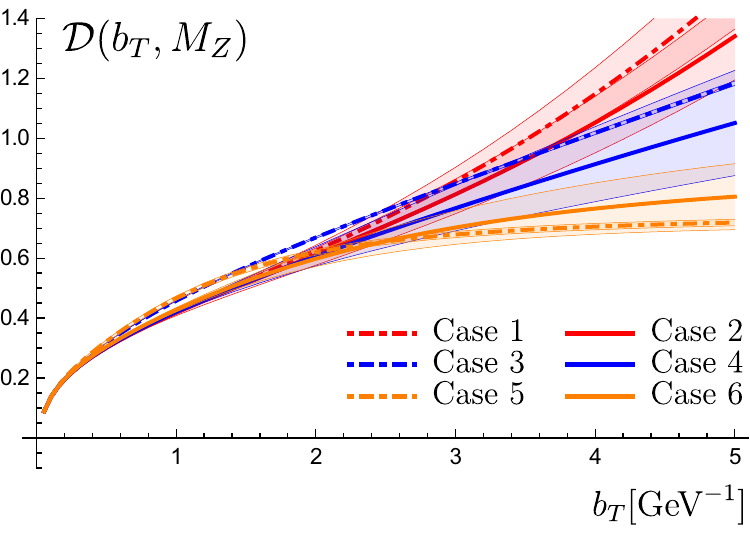}
\\
\includegraphics[width =0.3\textwidth]{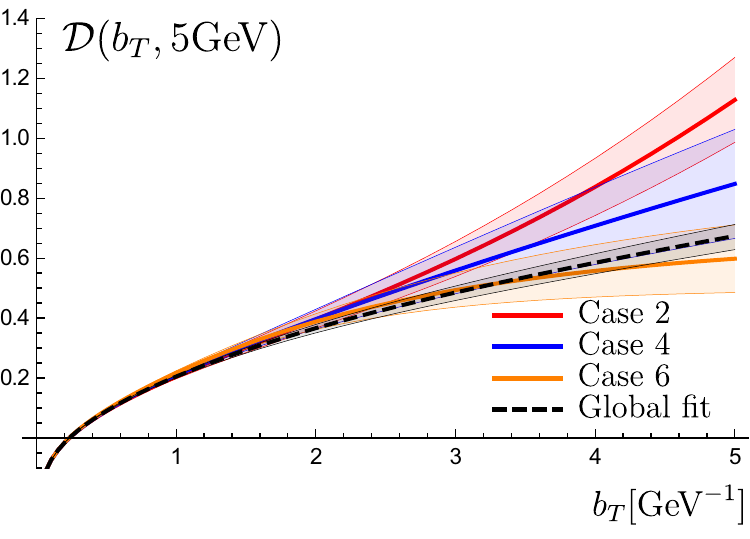}
\hspace{2cm}
\includegraphics[width =0.3\textwidth]{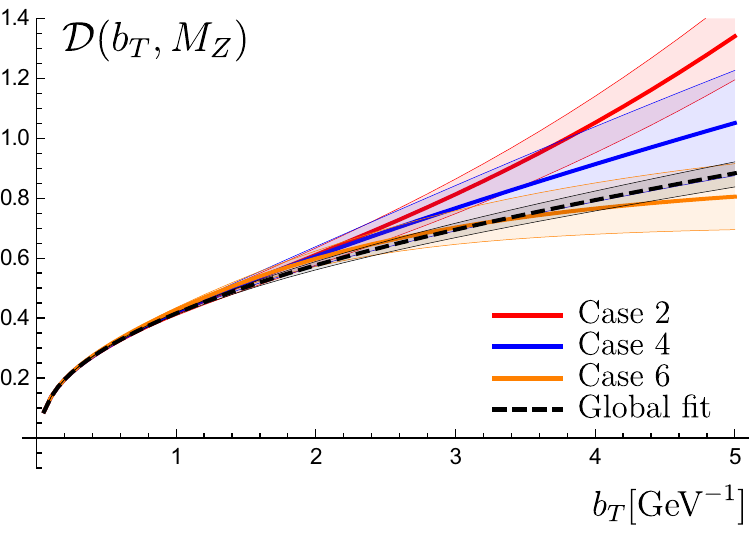}
\end{center}
\caption{ \textcolor{black}{Rapidity  evolution kernel 
 at $\mu=5$ GeV and  $\mu=M_Z$ GeV  for the different 
cases in Table~\ref{tab:case1}. In the lower panels  the result for  the global DY+SIDIS fit~\cite{Scimemi:2019cmh} is also plotted.
}
\label{fig:D}
}
\end{figure}

We explore the above point, associated with  correlations between 
$D_{\rm NP}$ and $f_{\rm NP}$,  by analyzing the  $  \vec b  $
dependence of   the rapidity evolution kernel $  \mathcal{D}(\mu,{\vec b})  $    in Fig.~\ref{fig:D}. We plot 
results for ${\cal D}$    from the different cases in Table~\ref{tab:case1},   at $\mu = M_Z$  
and $\mu = 5$ GeV.  Consider first the upper right panel  ($\mu = M_Z$). 
The two red curves  correspond to the  nonperturbative 
quadratic model in Eq.~(\ref{model:exp-gk}). The solid red curve is the result of  the 3-parameter fit in Table~\ref{tab:case1} (case 2), while   
the dashed red curve is the 
result of the 2-parameter fit in Table~\ref{tab:case1} (case 1). Similarly, the two yellow curves  correspond to the 
saturating model in Eq.~(\ref{model:exp-gk*})  (solid yellow  is the 3-parameter fit, while dashed yellow is 2-parameter), and 
the two blue curves  correspond to the linear  model  in   Eq.~(\ref{model:exp})   
 (solid blue  is the 3-parameter fit, while dashed blue is 2-parameter). 
 
 For  each of the three modeled 
  large-distance behaviors of  $  \mathcal{D}(\mu,\vec b)  $, the difference between the solid and  dashed curves in the upper right panel of 
  Fig.~\ref{fig:D}  measures the correlation between  the $D_{\rm NP}$ and $f_{\rm NP}$ nonperturbative effects, namely, it measures the impact of  
  the intrinsic $k_T$  on the determination of the rapidity evolution kernel. We see that in each case this impact is non-negligible. 
  If we look at the analogous results for lower masses in the  upper left panel  ($\mu = 5$ GeV), we see that for the 
  quadratic model particularly  (red curves) the impact of   intrinsic $k_T$ increases (even exceeding the uncertainty bands). 
  That is, although the quality of the fit from the quadratic model  is shown in Table~\ref{tab:case1} (case 2) to be comparable to that of 
  the saturating and linear models,  the quadratic model requires  a much more pronounced dependence  than the others 
  on the  intrinsic $k_T$   distribution, which  is revealed especially at low masses. 

Apart from the intrinsic $k_T$  correlations,  the differences among the three solid curves in the upper panels of 
Fig.~\ref{fig:D}      illustrate the current status  in the 
determination  of the large-$ | \vec b | $  behavior of the non-perturbative rapidity evolution kernel  from   fits to experimental data. 
As expected, the sensitivity of current LHC measurements to the long-distance region is limited, which results into 
sizeable uncertainty bands at large $ | \vec b | $. This sensitivity could be enhanced by precision measurements of the low-$ q_T$ DY  
spectrum at the LHC, with fine 
binning in $ q_T$,  for low masses $\mu \ll M_Z$ (see e.g. first results from LHCb~\cite{LHCb-CONF-2012-013}). 

For comparison, in the lower panels  of  Fig.~\ref{fig:D} 
 we also report the result for   ${\mathcal D}$ which is obtained  from the global   fit  to  
 Drell-Yan and semi-inclusive deep inelastic scattering data~\cite{Scimemi:2019cmh}  
  (grey curves in the two lower panels of Fig.~\ref{fig:D}).  The global fit   includes, besides  LHC data, also  
 data from low-energy experiments.   This fit is performed assuming the 
 linear model in  Eq.~(\ref{model:exp}).  It is interesting to observe that  the grey curves at $\mu = M_Z$  
and $\mu = 5$ GeV  in the lower  panels, compared to the blue curves  obtained from the  same linear 
model, are lower and  closer to the yellow curves (saturating behavior), reflecting   the role 
 of low-energy data in determining  long-distance features of ${\mathcal D}$.

\section{Conclusion} 

 Transverse momentum spectra in DY lepton pair production 
have been measured   at  the LHC  and at  lower-energy collider and fixed-target experiments.  
The  low-$q_T$  end of DY spectra is important 
for the extraction of the $W$-boson mass  and for hadron structure investigations. 

In this paper we have carried out a study of low-$q_T$ DY spectra based on the 
TMD factorization approach in Eq.~(\ref{def:xSec}), 
using the $\zeta$ prescription (\ref{th:zetaNP}) to treat the double scale evolution in 
Eqs.~(\ref{rapevolF}),(\ref{mass-evol}). 
This approach contains the perturbative TMD resummation through the 
coefficients in Table~\ref{tab:pert} and the non-perturbative  TMD contributions 
through $f_{\rm NP}$  (intrinsic $k_T$) and 
$D_{\rm NP}$ (non-perturbative Sudakov) in 
Eqs.~(\ref{model:TMDPDF}) and (\ref{model:rad}) 
(besides the non-perturbative collinear PDFs in Eq.~(\ref{model:TMDPDF})). 
As such,  it can be contrasted 
 with other approaches 
in the literature: on one hand, low-energy approaches based on 
fixed-scale parton model  which  include 
non-perturbative  TMD contributions  but 
do not include any perturbative resummation and/or evolution 
of TMDs; on the other hand,  high-energy approaches based on purely perturbative 
resummation and non-perturbative collinear PDFs, which do not include any 
non-perturbative  TMD contributions. 
We have  limited ourselves to considering 
the low-$q_T$ region  $ q_T \ll Q$, and not addressed issues of matching with finite-order perturbative corrections 
which are essential to treat the  region  $ q_T \sim Q$ (see e.g.~\cite{Camarda:2019zyx,Bizon:2019zgf,Martinez:2020fzs,Martinez:2019mwt}).

Using this theoretical framework,  we have performed 
fits to low-$q_T$ DY measurements 
from the LHC and from lower-energy experiments. The ultimate 
goal of these fits is to extract 
universal (non-perturbative) TMD distributions to be used in factorization formulas of 
the type  (\ref{def:xSec}), much in the spirit of the approaches discussed 
in~\cite{Aybat:2011zv,Hautmann:2014kza,Collins:2014jpa}. This will be essential to 
bring the use of TMDs for phenomenological analyses on a similar level as 
that of ordinary parton distributions. 
The determination of non-perturbative TMDs from fits to experimental measurements 
 is complementary to determinations from lattice QCD --- see e.g. 
 ongoing lattice  studies of $D_{\rm NP}$~\cite{Ebert:2019tvc,Ebert:2018gzl}. 
In this work we have focused on studying the sensitivity 
of LHC and lower-energy 
DY experiments to non-perturbative $f_{\rm NP}$   and 
$D_{\rm NP}$  contributions, and examining  their correlations with 
different extractions of collinear PDFs. 
To this end, we have 
defined model scenarios for $D_{\rm NP}$ in Eqs.~(\ref{model:exp-gk})-(\ref{model:exp-gk*}) 
 and $f_{\rm NP}$ in Eq.~(\ref{model:our_fNP}). 

 We have presented results from global DY fits (Figs.~\ref{fig:PDFs},\ref{fig:correlations} and Table~\ref{tab:result-new}) and 
 from LHC fits (Table~\ref{tab:case1}  and Fig.~\ref{fig:NPs}). These results indicate that,  
 while the strongest  sensitivity to the intrinsic  $k_T$ is provided by the low-energy  data, 
neglecting any intrinsic  $k_T$  at the LHC worsens the description  of  the lowest  $q_T$ bins in the DY spectrum, giving 
 higher $\chi^2$  values in the fit (see differences between cases 1 and 2, between cases 3 and 4, and between cases 5 and 6 in 
 Table~\ref{tab:case1}),  and causes a potential bias in the determination of 
 the rapidity evolution kernel $  \mathcal{D}(\mu,{\vec b})  $ 
 (see differences between cases 1 and 2, between cases 3 and 4, and between cases 5 and 6 in 
    Fig.~\ref{fig:D}).  A quantitative measure of the size of non-perturbative TMD effects is provided in Fig.~\ref{fig:NPs}  and 
   compared with perturbative theoretical uncertainties estimated from scale variations. Given the strong reduction of  these  uncertainties 
   achieved through the high logarithmic accuracy of perturbative resummations and the use of 
   the $\zeta $ prescription for scale-setting,   the residual uncertainty due to non-perturbative TMD effects is found to play a non-negligible 
   role for the    DY  spectrum  at the LHC   in the low-$q_T$ region, which increases  with decreasing DY masses.   

On the other hand, we see from the comparison of cases 2, 4 and 6 in  Fig.~\ref{fig:D} that the 
large-$| \vec b |$ behavior of     $  \mathcal{D} $   
 is not yet  constrained at present by available data both  at low energy and at the LHC. 
We have investigated and contrasted the hypotheses of quadratic behavior, which has traditionally been considered by 
extrapolation  from the perturbative result, and saturating behavior at long distances.   
We have observed in particular that the latter, besides being consistent  with current LHC fits, is also compatible with the result of a 
global fit based on an intermediate linear model, but including low-energy DY and SIDIS data. 
Given   the extraordinarily high experimental accuracy achieved in DY processes at the LHC,  this opens new opportunities for 
future LHC analyses. Specifically, extending   measurements of  the DY transverse momentum $q_T$, for low $q_T \ll Q$ 
and  with fine binning $ \leq  1 $~GeV,  into  the so far unexplored region of low masses $ Q <  40 $ GeV   will provide valuable 
new information on   $  \mathcal{D} $ at large $| \vec b |$, and thus enable  improved determinations of TMDs.

\vskip 0.9 cm 

\noindent {\bf Acknowledgments}.   We thank the participants of the ``Resummation, Evolution, Factorization" workshops in 
Krakow (November 2018) and Pavia (November 2019) for lively discussions on the topics of this work.  
F.H.  acknowledges the support and hospitality of DESY, Hamburg  while part of this work was being done.  
 I.S. is supported by the Spanish MECD grant FPA2016-75654-C2-2-P. 
 A.V. is supported by DFG grant N.430824754 as a part of the Research Unit FOR 2926.
 This project has received funding from 
 the European Union Horizon 2020 research and innovation program under grant agreement No 824093 (STRONG-2020).


 \bibliographystyle{JHEP}  
\bibliography{TMDref27feb}

\providecommand{\href}[2]{#2}\begingroup\raggedright\begin{thebibliography}{10}

\bibitem{Collins:1989gx}
J.~C. Collins, D.~E. Soper and G.~F. Sterman, \emph{{Factorization of Hard
  Processes in QCD}},
  \href{http://dx.doi.org/10.1142/9789814503266_0001}{\emph{Adv. Ser. Direct.
  High Energy Phys.} {\bfseries 5} (1989) 1--91},
  [\href{https://arxiv.org/abs/hep-ph/0409313}{{\ttfamily hep-ph/0409313}}].

\bibitem{Dokshitzer:1978hw}
Y.~L. Dokshitzer, D.~Diakonov and S.~I. Troian, \emph{{Hard Processes in
  Quantum Chromodynamics}},
  \href{http://dx.doi.org/10.1016/0370-1573(80)90043-5}{\emph{Phys. Rept.}
  {\bfseries 58} (1980) 269--395}.

\bibitem{Parisi:1979se}
G.~Parisi and R.~Petronzio, \emph{{Small Transverse Momentum Distributions in
  Hard Processes}},
  \href{http://dx.doi.org/10.1016/0550-3213(79)90040-3}{\emph{Nucl. Phys.}
  {\bfseries B154} (1979) 427--440}.

\bibitem{Curci:1979bg}
G.~Curci, M.~Greco and Y.~Srivastava, \emph{{{QCD} Jets From Coherent States}},
  \href{http://dx.doi.org/10.1016/0550-3213(79)90345-6}{\emph{Nucl. Phys.}
  {\bfseries B159} (1979) 451--468}.

\bibitem{Collins:1981uk}
J.~C. Collins and D.~E. Soper, \emph{{Back-To-Back Jets in QCD}},
  \href{http://dx.doi.org/10.1016/0550-3213(81)90339-4}{\emph{Nucl. Phys.}
  {\bfseries B193} (1981) 381}.

\bibitem{Collins:1984kg}
J.~C. Collins, D.~E. Soper and G.~F. Sterman, \emph{{Transverse Momentum
  Distribution in Drell-Yan Pair and W and Z Boson Production}},
  \href{http://dx.doi.org/10.1016/0550-3213(85)90479-1}{\emph{Nucl. Phys.}
  {\bfseries B250} (1985) 199--224}.

\bibitem{Vogt:2018miu}
A.~Vogt, F.~Herzog, S.~Moch, B.~Ruijl, T.~Ueda and J.~A.~M. Vermaseren,
  \emph{{Anomalous dimensions and splitting functions beyond the
  next-to-next-to-leading order}},
  \href{http://dx.doi.org/10.22323/1.303.0050}{\emph{PoS} {\bfseries LL2018}
  (2018) 050}, [\href{https://arxiv.org/abs/1808.08981}{{\ttfamily
  1808.08981}}].

\bibitem{Luo:2019szz}
M.-x. Luo, T.-Z. Yang, H.~X. Zhu and Y.~J. Zhu, \emph{{Quark Transverse Parton
  Distribution at the Next-to-Next-to-Next-to-Leading Order}},
  \href{https://arxiv.org/abs/1912.05778}{{\ttfamily 1912.05778}}.

\bibitem{Angeles-Martinez:2015sea}
R.~Angeles-Martinez et~al., \emph{{Transverse Momentum Dependent (TMD) parton
  distribution functions: status and prospects}},
  \href{http://dx.doi.org/10.5506/APhysPolB.46.2501}{\emph{Acta Phys. Polon.}
  {\bfseries B46} (2015) 2501--2534},
  [\href{https://arxiv.org/abs/1507.05267}{{\ttfamily 1507.05267}}].

\bibitem{Collins:1999dz}
J.~C. Collins and F.~Hautmann, \emph{{Infrared divergences and nonlightlike
  eikonal lines in Sudakov processes}},
  \href{http://dx.doi.org/10.1016/S0370-2693(99)01384-2}{\emph{Phys. Lett.}
  {\bfseries B472} (2000) 129--134},
  [\href{https://arxiv.org/abs/hep-ph/9908467}{{\ttfamily hep-ph/9908467}}].

\bibitem{Collins:2003fm}
J.~C. Collins, \emph{{What exactly is a parton density?}}, {\emph{Acta Phys.
  Polon.} {\bfseries B34} (2003) 3103},
  [\href{https://arxiv.org/abs/hep-ph/0304122}{{\ttfamily hep-ph/0304122}}].

\bibitem{Hautmann:2007uw}
F.~Hautmann, \emph{{Endpoint singularities in unintegrated parton
  distributions}},
  \href{http://dx.doi.org/10.1016/j.physletb.2007.08.081}{\emph{Phys. Lett.}
  {\bfseries B655} (2007) 26--31},
  [\href{https://arxiv.org/abs/hep-ph/0702196}{{\ttfamily hep-ph/0702196}}].

\bibitem{Collins:2011zzd}
J.~Collins, \emph{{Foundations of perturbative QCD}}.
\newblock Cambridge University Press, 2013.

\bibitem{GarciaEchevarria:2011rb}
M.~G. Echevarria, A.~Idilbi and I.~Scimemi, \emph{{Factorization Theorem For
  Drell-Yan At Low $q_T$ And Transverse Momentum Distributions
  On-The-Light-Cone}},
  \href{http://dx.doi.org/10.1007/JHEP07(2012)002}{\emph{JHEP} {\bfseries 07}
  (2012) 002}, [\href{https://arxiv.org/abs/1111.4996}{{\ttfamily 1111.4996}}].

\bibitem{Scimemi:2017etj}
I.~Scimemi and A.~Vladimirov, \emph{{Analysis of vector boson production within
  TMD factorization}},
  \href{http://dx.doi.org/10.1140/epjc/s10052-018-5557-y}{\emph{Eur. Phys. J.}
  {\bfseries C78} (2018) 89},
  [\href{https://arxiv.org/abs/1706.01473}{{\ttfamily 1706.01473}}].

\bibitem{Scimemi:2018xaf}
I.~Scimemi and A.~Vladimirov, \emph{{Systematic analysis of double-scale
  evolution}}, \href{http://dx.doi.org/10.1007/JHEP08(2018)003}{\emph{JHEP}
  {\bfseries 08} (2018) 003},
  [\href{https://arxiv.org/abs/1803.11089}{{\ttfamily 1803.11089}}].

\bibitem{Bertone:2019nxa}
V.~Bertone, I.~Scimemi and A.~Vladimirov, \emph{{Extraction of unpolarized
  quark transverse momentum dependent parton distributions from
  Drell-Yan/Z-boson production}},
  \href{http://dx.doi.org/10.1007/JHEP06(2019)028}{\emph{JHEP} {\bfseries 06}
  (2019) 028}, [\href{https://arxiv.org/abs/1902.08474}{{\ttfamily
  1902.08474}}].

\bibitem{Vladimirov:2019bfa}
A.~Vladimirov, \emph{{Pion-induced Drell-Yan processes within TMD
  factorization}}, \href{http://dx.doi.org/10.1007/JHEP10(2019)090}{\emph{JHEP}
  {\bfseries 10} (2019) 090},
  [\href{https://arxiv.org/abs/1907.10356}{{\ttfamily 1907.10356}}].

\bibitem{Scimemi:2016ffw}
I.~Scimemi and A.~Vladimirov, \emph{{Power corrections and renormalons in
  Transverse Momentum Distributions}},
  \href{http://dx.doi.org/10.1007/JHEP03(2017)002}{\emph{JHEP} {\bfseries 03}
  (2017) 002}, [\href{https://arxiv.org/abs/1609.06047}{{\ttfamily
  1609.06047}}].

\bibitem{Scimemi:2019cmh}
I.~Scimemi and A.~Vladimirov, \emph{{Non-perturbative structure of
  semi-inclusive deep-inelastic and Drell-Yan scattering at small transverse
  momentum}},  \href{https://arxiv.org/abs/1912.06532}{{\ttfamily 1912.06532}}.

\bibitem{Bacchetta:2019sam}
A.~Bacchetta, V.~Bertone, C.~Bissolotti, G.~Bozzi, F.~Delcarro, F.~Piacenza
  et~al., \emph{{Transverse-momentum-dependent parton distributions up to
  N$^3$LL from Drell-Yan data}},
  \href{https://arxiv.org/abs/1912.07550}{{\ttfamily 1912.07550}}.

\bibitem{Moch:2005tm}
S.~Moch, J.~A.~M. Vermaseren and A.~Vogt, \emph{{Three-loop results for quark
  and gluon form-factors}},
  \href{http://dx.doi.org/10.1016/j.physletb.2005.08.067}{\emph{Phys. Lett.}
  {\bfseries B625} (2005) 245--252},
  [\href{https://arxiv.org/abs/hep-ph/0508055}{{\ttfamily hep-ph/0508055}}].

\bibitem{Baikov:2009bg}
P.~A. Baikov, K.~G. Chetyrkin, A.~V. Smirnov, V.~A. Smirnov and M.~Steinhauser,
  \emph{{Quark and gluon form factors to three loops}},
  \href{http://dx.doi.org/10.1103/PhysRevLett.102.212002}{\emph{Phys. Rev.
  Lett.} {\bfseries 102} (2009) 212002},
  [\href{https://arxiv.org/abs/0902.3519}{{\ttfamily 0902.3519}}].

\bibitem{Vladimirov:2016dll}
A.~A. Vladimirov, \emph{{Soft-/rapidity- anomalous dimensions correspondence}},
  \href{http://dx.doi.org/10.1103/PhysRevLett.118.062001}{\emph{Phys. Rev.
  Lett.} {\bfseries 118} (2017) 062001},
  [\href{https://arxiv.org/abs/1610.05791}{{\ttfamily 1610.05791}}].

\bibitem{Li:2016ctv}
Y.~Li and H.~X. Zhu, \emph{{Bootstrapping Rapidity Anomalous Dimensions for
  Transverse-Momentum Resummation}},
  \href{http://dx.doi.org/10.1103/PhysRevLett.118.022004}{\emph{Phys. Rev.
  Lett.} {\bfseries 118} (2017) 022004},
  [\href{https://arxiv.org/abs/1604.01404}{{\ttfamily 1604.01404}}].

\bibitem{Echevarria:2012pw}
M.~G. Echevarria, A.~Idilbi, A.~Schafer and I.~Scimemi,
  \emph{{Model-Independent Evolution of Transverse Momentum Dependent
  Distribution Functions (TMDs) at NNLL}},
  \href{http://dx.doi.org/10.1140/epjc/s10052-013-2636-y}{\emph{Eur. Phys. J.}
  {\bfseries C73} (2013) 2636},
  [\href{https://arxiv.org/abs/1208.1281}{{\ttfamily 1208.1281}}].

\bibitem{Ladinsky:1993zn}
G.~A. Ladinsky and C.~P. Yuan, \emph{{The Nonperturbative regime in QCD
  resummation for gauge boson production at hadron colliders}},
  \href{http://dx.doi.org/10.1103/PhysRevD.50.R4239}{\emph{Phys. Rev.}
  {\bfseries D50} (1994) R4239},
  [\href{https://arxiv.org/abs/hep-ph/9311341}{{\ttfamily hep-ph/9311341}}].

\bibitem{Landry:1999an}
F.~Landry, R.~Brock, G.~Ladinsky and C.~P. Yuan, \emph{{New fits for the
  nonperturbative parameters in the CSS resummation formalism}},
  \href{http://dx.doi.org/10.1103/PhysRevD.63.013004}{\emph{Phys. Rev.}
  {\bfseries D63} (2001) 013004},
  [\href{https://arxiv.org/abs/hep-ph/9905391}{{\ttfamily hep-ph/9905391}}].

\bibitem{Landry:2002ix}
F.~Landry, R.~Brock, P.~M. Nadolsky and C.~P. Yuan, \emph{{Tevatron Run-1 $Z$
  boson data and Collins-Soper-Sterman resummation formalism}},
  \href{http://dx.doi.org/10.1103/PhysRevD.67.073016}{\emph{Phys. Rev.}
  {\bfseries D67} (2003) 073016},
  [\href{https://arxiv.org/abs/hep-ph/0212159}{{\ttfamily hep-ph/0212159}}].

\bibitem{Konychev:2005iy}
A.~V. Konychev and P.~M. Nadolsky, \emph{{Universality of the
  Collins-Soper-Sterman nonperturbative function in gauge boson production}},
  \href{http://dx.doi.org/10.1016/j.physletb.2005.12.063}{\emph{Phys. Lett.}
  {\bfseries B633} (2006) 710--714},
  [\href{https://arxiv.org/abs/hep-ph/0506225}{{\ttfamily hep-ph/0506225}}].

\bibitem{Hautmann:2007cx}
F.~Hautmann and D.~E. Soper, \emph{{Parton distribution function for quarks in
  an s-channel approach}},
  \href{http://dx.doi.org/10.1103/PhysRevD.75.074020}{\emph{Phys. Rev.}
  {\bfseries D75} (2007) 074020},
  [\href{https://arxiv.org/abs/hep-ph/0702077}{{\ttfamily hep-ph/0702077}}].

\bibitem{DAlesio:2014mrz}
U.~D'Alesio, M.~G. Echevarria, S.~Melis and I.~Scimemi, \emph{{Non-perturbative
  QCD effects in $q_{T}$ spectra of Drell-Yan and Z-boson production}},
  \href{http://dx.doi.org/10.1007/JHEP11(2014)098}{\emph{JHEP} {\bfseries 11}
  (2014) 098}, [\href{https://arxiv.org/abs/1407.3311}{{\ttfamily 1407.3311}}].

\bibitem{Aad:2014xaa}
{\scshape ATLAS} collaboration, G.~Aad et~al., \emph{{Measurement of the
  $Z/\gamma^*$ boson transverse momentum distribution in $pp$ collisions at
  $\sqrt{s}$ = 7 TeV with the ATLAS detector}},
  \href{http://dx.doi.org/10.1007/JHEP09(2014)145}{\emph{JHEP} {\bfseries 09}
  (2014) 145}, [\href{https://arxiv.org/abs/1406.3660}{{\ttfamily 1406.3660}}].

\bibitem{Aad:2015auj}
{\scshape ATLAS} collaboration, G.~Aad et~al., \emph{{Measurement of the
  transverse momentum and $\phi ^*_{\eta }$ distributions of Drell–Yan lepton
  pairs in proton–proton collisions at $\sqrt{s}=8$ TeV with the ATLAS
  detector}},
  \href{http://dx.doi.org/10.1140/epjc/s10052-016-4070-4}{\emph{Eur. Phys. J.}
  {\bfseries C76} (2016) 291},
  [\href{https://arxiv.org/abs/1512.02192}{{\ttfamily 1512.02192}}].

\bibitem{Chatrchyan:2011wt}
{\scshape CMS} collaboration, S.~Chatrchyan et~al., \emph{{Measurement of the
  Rapidity and Transverse Momentum Distributions of $Z$ Bosons in $pp$
  Collisions at $\sqrt{s}=7$ TeV}},
  \href{http://dx.doi.org/10.1103/PhysRevD.85.032002}{\emph{Phys. Rev.}
  {\bfseries D85} (2012) 032002},
  [\href{https://arxiv.org/abs/1110.4973}{{\ttfamily 1110.4973}}].

\bibitem{Khachatryan:2016nbe}
{\scshape CMS} collaboration, V.~Khachatryan et~al., \emph{{Measurement of the
  transverse momentum spectra of weak vector bosons produced in proton-proton
  collisions at $\sqrt{s}$ = 8 TeV}}, {\emph{Submitted to: JHEP} (2016) },
  [\href{https://arxiv.org/abs/1606.05864}{{\ttfamily 1606.05864}}].

\bibitem{Aaij:2015gna}
{\scshape LHCb} collaboration, R.~Aaij et~al., \emph{{Measurement of the
  forward $Z$ boson production cross-section in $pp$ collisions at $\sqrt{s}=7$
  TeV}}, \href{http://dx.doi.org/10.1007/JHEP08(2015)039}{\emph{JHEP}
  {\bfseries 08} (2015) 039},
  [\href{https://arxiv.org/abs/1505.07024}{{\ttfamily 1505.07024}}].

\bibitem{Aaij:2015zlq}
{\scshape LHCb} collaboration, R.~Aaij et~al., \emph{{Measurement of forward W
  and Z boson production in $pp$ collisions at $ \sqrt{s}=8 $ TeV}},
  \href{http://dx.doi.org/10.1007/JHEP01(2016)155}{\emph{JHEP} {\bfseries 01}
  (2016) 155}, [\href{https://arxiv.org/abs/1511.08039}{{\ttfamily
  1511.08039}}].

\bibitem{Aaij:2016mgv}
{\scshape LHCb} collaboration, R.~Aaij et~al., \emph{{Measurement of the
  forward Z boson production cross-section in pp collisions at $\sqrt{s} = 13$
  TeV}}, \href{http://dx.doi.org/10.1007/JHEP09(2016)136}{\emph{JHEP}
  {\bfseries 09} (2016) 136},
  [\href{https://arxiv.org/abs/1607.06495}{{\ttfamily 1607.06495}}].

\bibitem{Aidala:2018ajl}
{\scshape PHENIX} collaboration, C.~Aidala et~al., \emph{{Measurements of
  $\mu\mu$ pairs from open heavy flavor and Drell-Yan in $p+p$ collisions at
  $\sqrt{s}=200$ GeV}}, {\emph{Submitted to: Phys. Rev. D} (2018) },
  [\href{https://arxiv.org/abs/1805.02448}{{\ttfamily 1805.02448}}].

\bibitem{Aaltonen:2012fi}
{\scshape CDF} collaboration, T.~Aaltonen et~al., \emph{{Transverse momentum
  cross section of $e^+e^-$ pairs in the $Z$-boson region from $p\bar{p}$
  collisions at $\sqrt{s}=1.96$ TeV}},
  \href{http://dx.doi.org/10.1103/PhysRevD.86.052010}{\emph{Phys. Rev.}
  {\bfseries D86} (2012) 052010},
  [\href{https://arxiv.org/abs/1207.7138}{{\ttfamily 1207.7138}}].

\bibitem{Affolder:1999jh}
{\scshape CDF} collaboration, T.~Affolder et~al., \emph{{The transverse
  momentum and total cross section of $e^+e^-$ pairs in the $Z$ boson region
  from $p\bar{p}$ collisions at $\sqrt{s} = 1.8$ TeV}},
  \href{http://dx.doi.org/10.1103/PhysRevLett.84.845}{\emph{Phys. Rev. Lett.}
  {\bfseries 84} (2000) 845--850},
  [\href{https://arxiv.org/abs/hep-ex/0001021}{{\ttfamily hep-ex/0001021}}].

\bibitem{Abazov:2010kn}
{\scshape D0} collaboration, V.~M. Abazov et~al., \emph{{Measurement of the
  normalized $Z/\gamma^* -> \mu^+\mu^-$ transverse momentum distribution in
  $p\bar{p}$ collisions at $\sqrt{s}=1.96$ TeV}},
  \href{http://dx.doi.org/10.1016/j.physletb.2010.09.012}{\emph{Phys. Lett.}
  {\bfseries B693} (2010) 522--530},
  [\href{https://arxiv.org/abs/1006.0618}{{\ttfamily 1006.0618}}].

\bibitem{Abazov:2007ac}
{\scshape D0} collaboration, V.~M. Abazov et~al., \emph{{Measurement of the
  shape of the boson transverse momentum distribution in $p \bar{p} \to Z /
  \gamma^{*} \to e^+ e^- + X$ events produced at $\sqrt{s}$=1.96-TeV}},
  \href{http://dx.doi.org/10.1103/PhysRevLett.100.102002}{\emph{Phys. Rev.
  Lett.} {\bfseries 100} (2008) 102002},
  [\href{https://arxiv.org/abs/0712.0803}{{\ttfamily 0712.0803}}].

\bibitem{Abbott:1999wk}
{\scshape D0} collaboration, B.~Abbott et~al., \emph{{Measurement of the
  inclusive differential cross section for $Z$ bosons as a function of
  transverse momentum in $\bar{p}p$ collisions at $\sqrt{s} = 1.8$ TeV}},
  \href{http://dx.doi.org/10.1103/PhysRevD.61.032004}{\emph{Phys. Rev.}
  {\bfseries D61} (2000) 032004},
  [\href{https://arxiv.org/abs/hep-ex/9907009}{{\ttfamily hep-ex/9907009}}].

\bibitem{McGaughey:1994dx}
{\scshape E772} collaboration, P.~L. McGaughey et~al., \emph{{Cross-sections
  for the production of high mass muon pairs from 800-GeV proton bombardment of
  H-2}}, \href{http://dx.doi.org/10.1103/PhysRevD.50.3038,
  10.1103/PhysRevD.60.119903}{\emph{Phys. Rev.} {\bfseries D50} (1994)
  3038--3045}.

\bibitem{Moreno:1990sf}
G.~Moreno et~al., \emph{{Dimuon production in proton - copper collisions at
  $\sqrt{s}$ = 38.8-GeV}},
  \href{http://dx.doi.org/10.1103/PhysRevD.43.2815}{\emph{Phys. Rev.}
  {\bfseries D43} (1991) 2815--2836}.

\bibitem{Ito:1980ev}
A.~S. Ito et~al., \emph{{Measurement of the Continuum of Dimuons Produced in
  High-Energy Proton - Nucleus Collisions}},
  \href{http://dx.doi.org/10.1103/PhysRevD.23.604}{\emph{Phys. Rev.} {\bfseries
  D23} (1981) 604--633}.

\bibitem{web}
``\texttt{artemide} web-page, https://teorica.fis.ucm.es/artemide/ \\
  \texttt{artemide} repository,
  https://github.com/vladimirovalexey/artemide-public.''

\bibitem{Aghasyan:2017ctw}
{\scshape COMPASS} collaboration, M.~Aghasyan et~al.,
  \emph{{Transverse-momentum-dependent Multiplicities of Charged Hadrons in
  Muon-Deuteron Deep Inelastic Scattering}},
  \href{http://dx.doi.org/10.1103/PhysRevD.97.032006}{\emph{Phys. Rev.}
  {\bfseries D97} (2018) 032006},
  [\href{https://arxiv.org/abs/1709.07374}{{\ttfamily 1709.07374}}].

\bibitem{Airapetian:2012ki}
{\scshape HERMES} collaboration, A.~Airapetian et~al., \emph{{Multiplicities of
  charged pions and kaons from semi-inclusive deep-inelastic scattering by the
  proton and the deuteron}},
  \href{http://dx.doi.org/10.1103/PhysRevD.87.074029}{\emph{Phys. Rev.}
  {\bfseries D87} (2013) 074029},
  [\href{https://arxiv.org/abs/1212.5407}{{\ttfamily 1212.5407}}].

\bibitem{Camarda:2019zyx}
S.~Camarda et~al., \emph{{DYTurbo: Fast predictions for Drell-Yan processes}},
  \href{https://arxiv.org/abs/1910.07049}{{\ttfamily 1910.07049}}.

\bibitem{Catani:2015vma}
S.~Catani, D.~de~Florian, G.~Ferrera and M.~Grazzini, \emph{{Vector boson
  production at hadron colliders: transverse-momentum resummation and leptonic
  decay}}, \href{http://dx.doi.org/10.1007/JHEP12(2015)047}{\emph{JHEP}
  {\bfseries 12} (2015) 047},
  [\href{https://arxiv.org/abs/1507.06937}{{\ttfamily 1507.06937}}].

\bibitem{Bizon:2019zgf}
W.~Bizon, A.~Gehrmann-De~Ridder, T.~Gehrmann, N.~Glover, A.~Huss, P.~F. Monni
  et~al., \emph{{The transverse momentum spectrum of weak gauge bosons at N
  ${}^3$ LL + NNLO}},
  \href{http://dx.doi.org/10.1140/epjc/s10052-019-7324-0}{\emph{Eur. Phys. J.}
  {\bfseries C79} (2019) 868},
  [\href{https://arxiv.org/abs/1905.05171}{{\ttfamily 1905.05171}}].

\bibitem{Bizon:2018foh}
W.~Bizon, X.~Chen, A.~Gehrmann-De~Ridder, T.~Gehrmann, N.~Glover, A.~Huss
  et~al., \emph{{Fiducial distributions in Higgs and Drell-Yan production at
  N$^{3}$LL+NNLO}},
  \href{http://dx.doi.org/10.1007/JHEP12(2018)132}{\emph{JHEP} {\bfseries 12}
  (2018) 132}, [\href{https://arxiv.org/abs/1805.05916}{{\ttfamily
  1805.05916}}].

\bibitem{Martinez:2019mwt}
A.~Bermudez~Martinez et~al., \emph{{Production of Z-bosons in the parton
  branching method}},
  \href{http://dx.doi.org/10.1103/PhysRevD.100.074027}{\emph{Phys. Rev.}
  {\bfseries D100} (2019) 074027},
  [\href{https://arxiv.org/abs/1906.00919}{{\ttfamily 1906.00919}}].

\bibitem{Martinez:2018jxt}
A.~Bermudez~Martinez, P.~Connor, H.~Jung, A.~Lelek, R.~Zlebcik, F.~Hautmann
  et~al., \emph{{Collinear and TMD parton densities from fits to precision DIS
  measurements in the parton branching method}},
  \href{http://dx.doi.org/10.1103/PhysRevD.99.074008}{\emph{Phys. Rev.}
  {\bfseries D99} (2019) 074008},
  [\href{https://arxiv.org/abs/1804.11152}{{\ttfamily 1804.11152}}].

\bibitem{Martinez:2020fzs}
A.~B. Martinez et~al., \emph{{The transverse momentum spectrum of low mass
  Drell-Yan production at next-to-leading order in the parton branching
  method}},  \href{https://arxiv.org/abs/2001.06488}{{\ttfamily 2001.06488}}.

\bibitem{Bacchetta:2019tcu}
A.~Bacchetta, G.~Bozzi, M.~Lambertsen, F.~Piacenza, J.~Steiglechner and
  W.~Vogelsang, \emph{{Difficulties in the description of Drell-Yan processes
  at moderate invariant mass and high transverse momentum}},
  \href{http://dx.doi.org/10.1103/PhysRevD.100.014018}{\emph{Phys. Rev.}
  {\bfseries D100} (2019) 014018},
  [\href{https://arxiv.org/abs/1901.06916}{{\ttfamily 1901.06916}}].

\bibitem{Collins:2000gd}
J.~C. Collins and F.~Hautmann, \emph{{Soft gluons and gauge invariant
  subtractions in NLO parton shower Monte Carlo event generators}},
  \href{http://dx.doi.org/10.1088/1126-6708/2001/03/016}{\emph{JHEP} {\bfseries
  03} (2001) 016}, [\href{https://arxiv.org/abs/hep-ph/0009286}{{\ttfamily
  hep-ph/0009286}}].

\bibitem{Ball:2017nwa}
{\scshape NNPDF} collaboration, R.~D. Ball et~al., \emph{{Parton distributions
  from high-precision collider data}},
  \href{http://dx.doi.org/10.1140/epjc/s10052-017-5199-5}{\emph{Eur. Phys. J.}
  {\bfseries C77} (2017) 663},
  [\href{https://arxiv.org/abs/1706.00428}{{\ttfamily 1706.00428}}].

\bibitem{Abramowicz:2015mha}
{\scshape H1, ZEUS} collaboration, H.~Abramowicz et~al., \emph{{Combination of
  measurements of inclusive deep inelastic ${e^{\pm }p}$ scattering cross
  sections and QCD analysis of HERA data}},
  \href{http://dx.doi.org/10.1140/epjc/s10052-015-3710-4}{\emph{Eur. Phys. J.}
  {\bfseries C75} (2015) 580},
  [\href{https://arxiv.org/abs/1506.06042}{{\ttfamily 1506.06042}}].

\bibitem{Dulat:2015mca}
S.~Dulat, T.-J. Hou, J.~Gao, M.~Guzzi, J.~Huston, P.~Nadolsky et~al.,
  \emph{{New parton distribution functions from a global analysis of quantum
  chromodynamics}},
  \href{http://dx.doi.org/10.1103/PhysRevD.93.033006}{\emph{Phys. Rev.}
  {\bfseries D93} (2016) 033006},
  [\href{https://arxiv.org/abs/1506.07443}{{\ttfamily 1506.07443}}].

\bibitem{Harland-Lang:2015nxa}
L.~A. Harland-Lang, A.~D. Martin, P.~Motylinski and R.~S. Thorne,
  \emph{{Uncertainties on $\alpha _S$ in the MMHT2014 global PDF analysis and
  implications for SM predictions}},
  \href{http://dx.doi.org/10.1140/epjc/s10052-015-3630-3}{\emph{Eur. Phys. J.}
  {\bfseries C75} (2015) 435},
  [\href{https://arxiv.org/abs/1506.05682}{{\ttfamily 1506.05682}}].

\bibitem{Butterworth:2015oua}
J.~Butterworth et~al., \emph{{PDF4LHC recommendations for LHC Run II}},
  \href{http://dx.doi.org/10.1088/0954-3899/43/2/023001}{\emph{J. Phys.}
  {\bfseries G43} (2016) 023001},
  [\href{https://arxiv.org/abs/1510.03865}{{\ttfamily 1510.03865}}].

\bibitem{LHCb-CONF-2012-013}
{\scshape LHCb} collaboration, J.~Anderson and K.~Mueller, \emph{{Inclusive low
  mass Drell-Yan production in the forward region at $\sqrt{s}$ = 7 TeV,
  LHCb-CONF-2012-013}}, .

\bibitem{Aybat:2011zv}
S.~M. Aybat and T.~C. Rogers, \emph{{TMD Parton Distribution and Fragmentation
  Functions with QCD Evolution}},
  \href{http://dx.doi.org/10.1103/PhysRevD.83.114042}{\emph{Phys. Rev.}
  {\bfseries D83} (2011) 114042},
  [\href{https://arxiv.org/abs/1101.5057}{{\ttfamily 1101.5057}}].

\bibitem{Hautmann:2014kza}
F.~Hautmann, H.~Jung, M.~Kraemer, P.~J. Mulders, E.~R. Nocera, T.~C. Rogers
  et~al., \emph{{TMDlib and TMDplotter: library and plotting tools for
  transverse-momentum-dependent parton distributions}},
  \href{http://dx.doi.org/10.1140/epjc/s10052-014-3220-9}{\emph{Eur. Phys. J.}
  {\bfseries C74} (2014) 3220},
  [\href{https://arxiv.org/abs/1408.3015}{{\ttfamily 1408.3015}}].

\bibitem{Collins:2014jpa}
J.~Collins and T.~Rogers, \emph{{Understanding the large-distance behavior of
  transverse-momentum-dependent parton densities and the Collins-Soper
  evolution kernel}},
  \href{http://dx.doi.org/10.1103/PhysRevD.91.074020}{\emph{Phys. Rev.}
  {\bfseries D91} (2015) 074020},
  [\href{https://arxiv.org/abs/1412.3820}{{\ttfamily 1412.3820}}].

\bibitem{Ebert:2019tvc}
M.~A. Ebert, I.~W. Stewart and Y.~Zhao, \emph{{Renormalization and Matching for
  the Collins-Soper Kernel from Lattice QCD}},
  \href{https://arxiv.org/abs/1910.08569}{{\ttfamily 1910.08569}}.

\bibitem{Ebert:2018gzl}
M.~A. Ebert, I.~W. Stewart and Y.~Zhao, \emph{{Determining the Nonperturbative
  Collins-Soper Kernel From Lattice QCD}},
  \href{http://dx.doi.org/10.1103/PhysRevD.99.034505}{\emph{Phys. Rev.}
  {\bfseries D99} (2019) 034505},
  [\href{https://arxiv.org/abs/1811.00026}{{\ttfamily 1811.00026}}].

\end{thebibliography}\endgroup
\end{document}